\documentclass[aps,reprint,prb,amsmath,noeprint]{revtex4-2}
\usepackage{graphics}
\usepackage{natbib}
\usepackage{graphicx}
\usepackage{epsfig}
\usepackage{amsmath}
\usepackage{bm}
\usepackage{lineno}
\usepackage{color}
\usepackage{dcolumn}
\usepackage{xcolor}
\begin{document}
\title{The interplay of large two-magnon ferromagnetic resonance linewidths and low Gilbert damping in Heusler thin films}
\author{W.~K.~Peria,$^1$ T.~A.~Peterson,$^1$ A.~P.~McFadden,$^2$ T.~Qu,$^3$ C.~Liu,$^1$ C.~J.~Palmstr{\o}m,$^{2,4}$ and P.~A.~Crowell$^1$}
\affiliation{$^1$School of Physics and Astronomy, University of Minnesota, Minneapolis, Minnesota 55455\\
$^2$Department of Electrical \& Computer Engineering, University of California, Santa Barbara, California 93106\\
$^3$Department of Electrical and Computer Engineering, University of Minnesota, Minneapolis, Minnesota 55455\\
$^4$Department of Materials, University of California, Santa Barbara, California 93106}
\begin{abstract}
We report on broadband ferromagnetic resonance linewidth measurements performed on epitaxial Heusler thin films. A large and anisotropic two-magnon scattering linewidth broadening is observed for measurements with the magnetization lying in the film plane, while linewidth measurements with the magnetization saturated perpendicular to the sample plane reveal low Gilbert damping constants of $(1.5\pm0.1)\times 10^{-3}$, $(1.8\pm0.2)\times 10^{-3}$, and $<8\times 10^{-4}$ for Co$_2$MnSi/MgO, Co$_2$MnAl/MgO, and Co$_2$FeAl/MgO, respectively. The in-plane measurements are fit to a model combining Gilbert and two-magnon scattering contributions to the linewidth, revealing a characteristic disorder lengthscale of 10-100~nm.
\end{abstract}
\maketitle
\section{\label{intro}Introduction}
The theoretical understanding of the damping mechanism believed to govern longitudinal magnetization relaxation in metallic ferromagnets, originally due to Kambersk\'y \cite{Kambersky1970,Kambersky1976}, has in recent years resulted in quantitative damping estimates for realistic transition metal band structures \cite{Steiauf2005, Gilmore2007, Gilmore2010}. Although of great interest where engineering of damping is desired \cite{Ralph2008}, these calculations  remain largely uncompared to experimental data. Kambersk\'y damping may be characterized by the so-called Gilbert damping constant $\alpha$ in the Landau-Lifshitz-Gilbert macrospin torque equation of motion, and formally describes how the spin-orbit interaction in itinerant electron systems results in damping of magnetization dynamics \cite{Kambersky1976}. \citet{Schoen2016} have reported that $\alpha$ is minimized for Co-Fe alloy compositions at which the density-of-states at the Fermi level is minimized, in reasonable agreement with Kambersk\'y model predictions \cite{Mankovsky2013}. Furthermore, half-metallic, or near half-metallic ferromagnets such as full-Heusler compounds have been predicted to demonstrate an ultralow Kambersk\'y $\alpha$ ($\leq 10^{-3}$) due to their spin-resolved band structure near the Fermi level \cite{Liu2009}. Finally, anisotropy of the Kambersk\'y damping in single crystals has been predicted, which is more robust for Fermi surfaces with single-band character \cite{Gilmore2010,Qu2014}.

The Gilbert damping constant is often reported through measurements of the ferromagnetic resonance (FMR) linewidth $\Delta H$, which may be expressed as a sum of individual contributions
\begin{equation}\label{LWtotal}
\Delta H = \frac{2\alpha f}{\gamma}+\Delta H_0+\Delta H_{TMS},
\end{equation}
where the first term is the Gilbert damping linewidth ($f$ is the FMR frequency, $\gamma$ is the gyromagnetic ratio), $\Delta H_0$ is a frequency-independent inhomogeneous broadening, and $\Delta H_{TMS}$ represents an extrinsic two-magnon scattering (TMS) linewidth contribution \cite{Arias1999,McMichael2004} that is, in general, a nonlinear function of frequency. In recent years it has been realized that TMS linewidths are pervasive for the conventional in-plane geometry of thin film FMR measurements, requiring either the perpendicular-to-plane FMR geometry \cite{Schoen2015} (for which TMS processes are suppressed) or sufficiently broadband measurements \cite{Woltersdorf2004} to extract the bare Gilbert $\alpha$. For instance, recent FMR linewidth studies on Heusler compounds have reported distinct TMS linewidths \cite{Mizukami2009,Qiao2016}, which challenged simple inference of the Gilbert $\alpha$.

In this article, we present FMR linewidth measurements for epitaxial Heusler thin films for all principal orientations of the magnetization with respect to the symmetry axes. For the in-plane configuration, large and anisotropic TMS-dominated linewidths are observed.  In the perpendicular-to-plane configuration, for which the TMS process is inactive \cite{Arias1999}, the Gilbert $\alpha$ and inhomogeneous broadening are measured. We find evidence of a low ($\sim$10$^{-3}$) Gilbert $\alpha$ in these Heusler thin films, accompanied by a large and anisotropic TMS contribution to the linewdith for in-plane magnetization. We conclude by discussing the interplay of low Gilbert $\alpha$ and large TMS, and we emphasize the nature by which the TMS may conceal the presence of anisotropic Kambersk\'y $\alpha$.
\section{Samples}\label{samples}
The Heusler alloy films used for these measurements were grown by molecular beam epitaxy (MBE) by co-evaporation of elemental sources in ultrahigh vacuum (UHV). The MgO(001) substrates were annealed at 700~$^{\circ}$C in UHV followed by growth of a 20 nm thick MgO buffer layer by e-beam evaporation at a substrate temperature of 630~$^{\circ}$C. The 10 nm thick Co$_2$MnAl and Co$_2$MnSi films were grown on the MgO buffer layers at room temperature and then annealed at 600~$^{\circ}$C for 15 minutes \textit{in situ} in order to improve crystalline order and surface morphology. The 24 nm thick Co$_2$FeAl sample was grown using the same MgO substrate and buffer layer preparation, but at a substrate temperature of 250~$^{\circ}$C with no post-growth anneal. Reflection high energy electron diffraction (RHEED) was monitored during and after growth of all samples and confirmed the expected epitaxial relationship of MgO(001)$\langle$110$\rangle$ $\vert\vert$ Heusler(001)$\langle$100$\rangle$. X-ray diffraction (XRD) demonstrated the existence of a single phase of (001)-oriented Heusler, along with the presence of the (002) reflection, confirming at least B2 ordering in all cases. In addition, for the Co$_2$MnSi film only, the ($111$) reflection was observed, indicating L2$_1$ ordering [see Fig. \ref{fig:1}(a)]. All of the films were capped with several nm of e-beam evaporated AlOx for passivation prior to atmospheric exposure. The effective magnetization for the 24~nm thick Co$_2$FeAl film was determined from anomalous Hall effect saturation field to be 1200~emu/cm$^3$, which is consistent with measurements of Ref.\ \cite{Cui2014} for L2$_1$ or B2-ordered films, along with 990~emu/cm$^3$ and 930~emu/cm$^3$ for the Co$_2$MnSi and Co$_2$MnAl films, respectively. Hereafter, we will refer to the Co$_2$MnSi(10~nm)/MgO as the ``CMS" film, the Co$_2$MnAl(10~nm)/MgO film as the ``CMA" film, and the Co$_2$FeAl(24~nm)/MgO film as the ``CFA'' film.
\section{\label{sec:experiment}Experiment}
Broadband FMR linewidth measurements were performed at room temperature with a coplanar waveguide (CPW) transmission setup, similar to that discussed in detail in Refs.\ \cite{Kalarickal2006, Montoya2014}, placed between the pole faces of an electromagnet. A cleaved piece of the sample ($\sim$2~mm$\times$1~mm) was placed face-down over the centerline of the CPW. A rectifying diode was used to detect the transmitted microwave power, and a $\sim$100~Hz magnetic field modulation was used for lock-in detection of the transmitted power, resulting in a signal $\propto d\chi/dH$ (where $\chi$ is the film dynamic magnetic susceptibility). The excitation frequency could be varied from 0-50 GHz, and a microwave power near 0~dBm was typically used. It was verified that all measurements discussed in this article were in the small precession cone angle, linear regime. The orientation of the applied magnetic field could be rotated to arbitrary angle in the film plane (IP), or applied perpendicular to the film plane (PP). We emphasize again that TMS contributions are suppressed in the PP configuration \cite{McMichael2004}. The resonance fields were fit as a function of applied frequency in order to extract various magnetic properties of the films.

\begin{figure}
\includegraphics{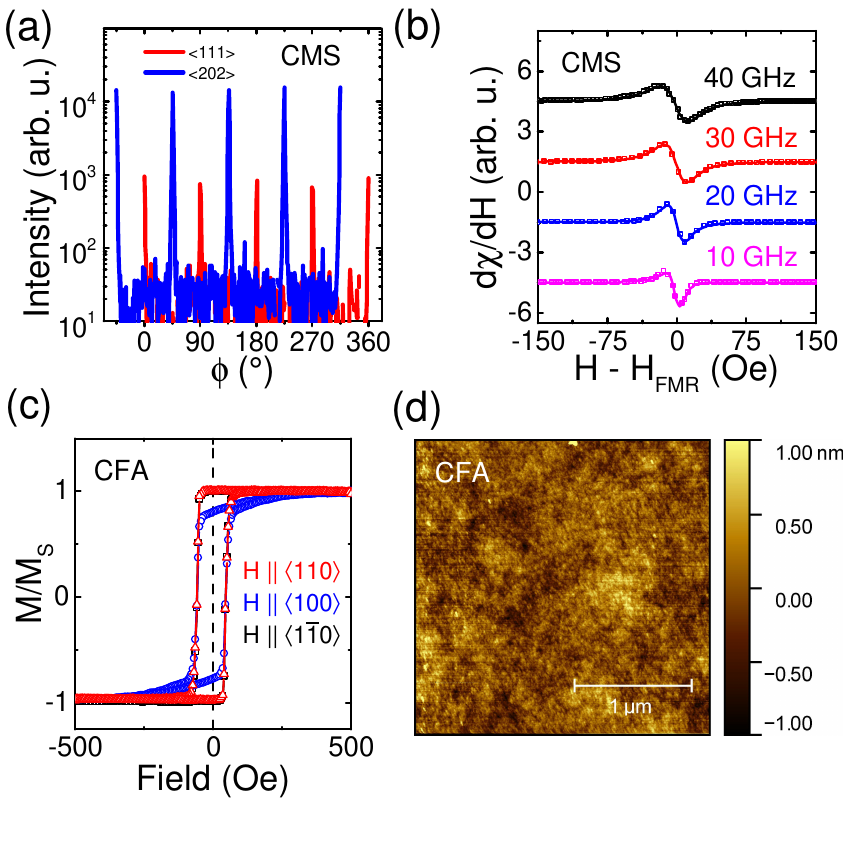}
\caption{(a) Wide-angle x-ray diffraction $\phi$-scans of $\langle202\rangle$ (blue) and $\langle111\rangle$ (red) peaks for the CMS film. (b) Typical derivative susceptibility lineshapes for these samples at different microwave excitation frequencies. The fits are shown as solid lines. (c) In-plane hysteresis loops for CFA obtained with a vibrating-sample magnetometer (VSM). (d) Atomic force microscopy (AFM) image of surface topography for CFA. RMS roughness is 0.2~nm.}
\label{fig:1}
\end{figure}

The magnetic free energy per unit volume used to generate the resonance conditions for these samples is given by
\begin{equation}\label{eq:FT}
  F_\mathbf{M} = - \mathbf{M} \cdot \mathbf{H} + K_1 \sin^2 \phi \cos^2 \phi + 2 \pi M_{eff}^2 \cos^2 \theta,
\end{equation}
where $\mathbf{H}$ is the applied field, $\phi$ and $\theta$ are the azimuthal and polar angles of the magnetization, respectively, $K_1$ is a first order in-plane cubic anisotropy constant, and $4 \pi M_{eff}$ is the PP saturation field, which includes the usual demagnetization energy and a first order uniaxial anisotropy due to interfacial effects. The parameters obtained by fitting to Eq.\ \ref{eq:FT} are shown in Table \ref{tab:magneticparameters}. The uncertainty in these parameters was estimated by measuring a range of different sample pieces, and using the standard deviation of the values as the error bar. The long-range inhomogeneity characteristic of epitaxial samples makes this a more accurate estimate of the uncertainty than the fitting error. The magnetic-field-swept FMR lineshapes were fit to the derivative of Lorentzian functions \cite{Montoya2014} in order to extract the full-width at half-maximum linewidths $\Delta H$ [magnetic field units, Fig.\ \ref{fig:1}(b)], which are the focus of this article.
\begin{table*}
    \caption{Summary of the magnetic properties extracted from the dependence of the resonance field on applied frequency for both field in-plane ($\vert\vert$) and field perpendicular-to-plane ($\perp$) configurations, along with the Gilbert $\alpha$ and inhomogeneous broadening from the perpendicular-to-plane configuration. $2 K_1/M_s$ and $4 \pi M_{eff}$ are the in-plane and perpendicular-to-plane anisotropy fields, respectively (see Eq.\ \ref{eq:FT}), and $g$ is the Land\'e $g$-factor.}
    \label{tab:magneticparameters}
    \begin{ruledtabular}
    \begin{tabular}{l c c c c c c c}
       \\[-9pt]
       Sample & $2 K_1/M_s$ (Oe)  & $4\pi M_{eff}^{\vert\vert}$ (kOe) & $4\pi M_{eff}^{\perp}$ (kOe) & $g^{\vert\vert}$ & $g^{\perp}$ & $\alpha_{001} (\times 10^{-3})$ & $\Delta H_{0}$ (Oe) \\[4pt]
        \hline
        \\[-6pt]
        CMS & 280 & 12.3 & 13.3 & 2.04 & 2.04 & $1.5 \pm 0.1$ & $9 \pm 1$\\
        CMA & 35 & 11.3 & 11.7 & 2.06 & 2.08 & $1.8 \pm 0.2$ & $12 \pm 3$\\
        CFA & 230 & 15.1 & 15.5 & 2.06 & 2.07 & $<$ 0.8 & $100 \pm 6$\\
        CFA 500~$^\circ$C anneal & N/A & N/A & 15.1 & N/A & 2.07 & $1.1 \pm 0.1$ & $45 \pm 1$
    \end{tabular}
    \end{ruledtabular}
\end{table*}
The maximum resonant frequency was determined by the maximum magnetic field that could be applied for both IP and PP electromagnet configurations, which was 10.6~kOe and 29~kOe, respectively. For the IP measurement, the angle of the applied field in the plane of the film was varied to determine the in-plane magnetocrystalline anisotropy of our samples, which was fourfold-symmetric for the three films characterized in this article. The anisotropy was confirmed using vibrating-sample magnetometry (VSM) measurements, an example of which is shown in Fig.\ \ref{fig:1}(c), which shows IP easy and hard axis hysteresis loops for the CFA film. For the PP measurement, alignment was verified to within $\sim$0.1$^{\circ}$ to ensure magnetization saturation just above the PP anisotropy field, thus minimizing field-dragging contributions to the linewidth.
\begin{figure}
\includegraphics{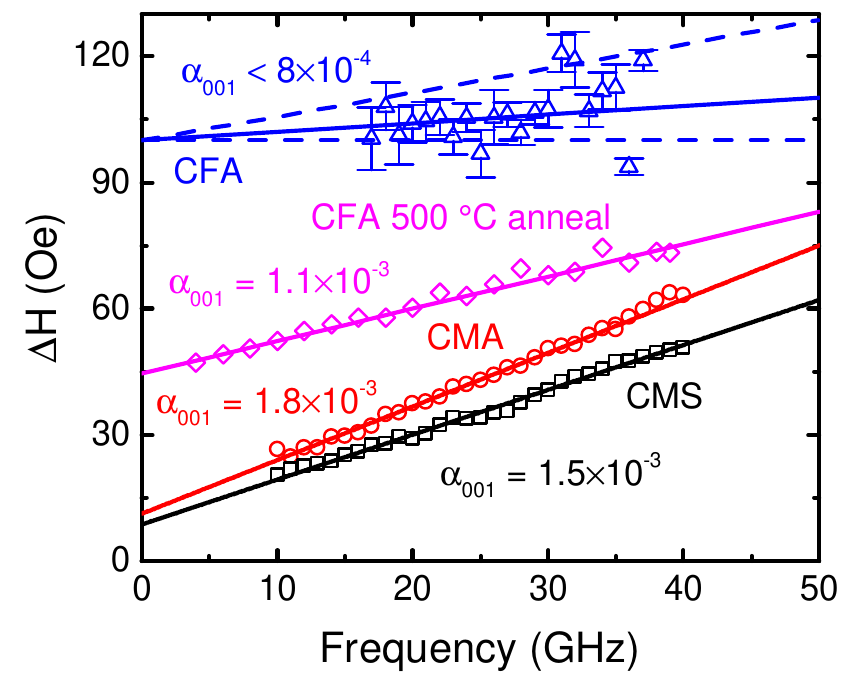}
\caption{Linewidths as a function of frequency with the field applied perpendicular to plane, for which two-magnon scattering is inactive.  The black squares are data for the CMS film, the red circles are for the CMA film, and the blue triangles are for the CFA film. In addition, linewidths are shown for a CFA film that was annealed at 500~$^{\circ}$C \textit{ex situ} (magenta diamonds). Corresponding linear fits are shown along with the extracted Gilbert damping factor $\alpha$. The blue dashed lines indicate an upper bound of $\alpha_{001} = 8\times 10^{-4}$ and a lower bound of $\alpha_{001} = 0$ for CFA.}
\label{fig:PP_summary}
\end{figure}
\section{Results and Analysis}\label{data}
\begin{figure}
\includegraphics{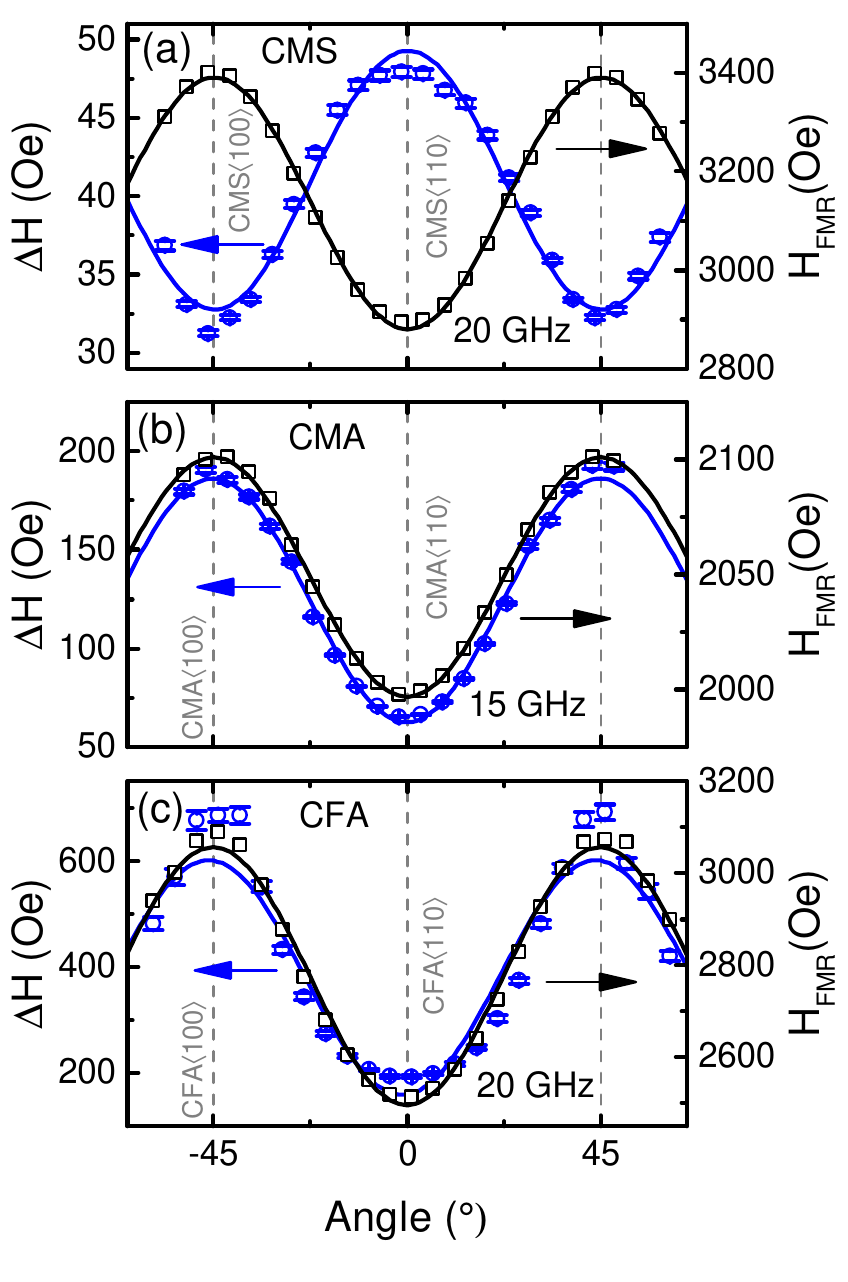}
\caption{Azimuthal angular dependence of the linewidths (left ordinate, blue circles) and resonance fields (right ordinate, black squares) for (a) CMS, (b) CMA, and (c) CFA. The excitation frequency was 20 GHz for CMS, 15 GHz for CMA, and 20 GHz for CFA. The solid lines are sinusoidal fits.}
\label{fig:angledependence}
\end{figure}
\subsection{\label{PP}Perpendicular-to-plane linewidths}
First we discuss the results of the PP measurement.  As stated in Sec. \ref{sec:experiment}, the TMS extrinsic broadening mechanism is suppressed when the magnetization is normal to the plane of the film. We can thus fit our data to Eq. \ref{LWtotal} with $\Delta H_{TMS} = 0$, greatly simplifying the extraction of the Gilbert damping constant $\alpha$ and the inhomogeneous broadening $\Delta H_0$. Prior knowledge of $\Delta H_0$ is particularly important for constraining the analysis of the IP measurements, as we shall discuss.

The dependence of $\Delta H$ on frequency for the CMS, CMA, and CFA films in the PP configuration is summarized in Fig.\ \ref{fig:PP_summary}, in which fits to Eq.\ \ref{LWtotal} are shown with $\Delta H_{TMS}$ set to zero. For the CMS film, $\alpha_{001} = (1.5\pm0.1)\times 10^{-3}$ and $\Delta H_0 = 9$~Oe, while for the CMA film $\alpha_{001} = (1.8\pm0.2)\times 10^{-3}$ and $\Delta H_0 = 12$~Oe. Co$_2$MnSi$_{2/3}$Al$_{1/3}$/MgO and Co$_2$MnSi$_{1/3}$Al$_{2/3}$/MgO films (both 10~nm thick) were also measured, with Gilbert damping values of $\alpha_{001} = (1.8\pm0.2)\times 10^{-3}$ and $\alpha_{001} = (1.5\pm0.1)\times 10^{-3}$, respectively (not shown). For CFA, we obtained a damping value of $\alpha_{001} = 3\times 10^{-4}$  with an upper bound of $\alpha_{001} < 8\times 10^{-4}$ and $\Delta H_0 = 100$~Oe. These fit parameters are also contained in Table \ref{tab:magneticparameters}. The source of the large inhomogeneous broadening for the CFA film is unclear: AFM measurements [Fig.\ \ref{fig:1}(d)] along with XRD indicate that the film is both crystalline and smooth. Note that the range of frequencies shown in Fig.\ \ref{fig:PP_summary} are largely governed by considerations involving the Kittel equation \cite{Kittel1948}: measurements below 10~GHz were not used due to the increasing influence of slight misalignment on $\Delta H$ (through field-dragging) for resonant fields just above the saturation value. A piece of the CFA sample was annealed at 500~$^{\circ}$C \textit{ex situ}, which reduced the inhomogenoeus broadening to $\sim$45 Oe (still a relatively large value) and increased the Gilbert damping to $\alpha_{001} = 1.1\times 10^{-3}$ (similar behavior in CFA was seen in Ref.\ \cite{Kumar2017}). The constraint of $\alpha_{001} < 8\times 10^{-4}$ is among the lowest of reported Gilbert damping constants for metallic ferromagnets, but the $\alpha\sim 10^{-4}$ range is not unexpected based on Kambersk\'y model calculations performed for similar full-Heusler compounds \cite{Liu2009} or other recent experimental reports \cite{Oogane2018,Guillemard2019}. It should be noted that \citet{Schoen2016} have recently reported $\alpha = 5\times 10^{-4}$ for Co$_{25}$Fe$_{75}$ thin films, where spin pumping and radiative damping contributions were subtracted from the raw measurement. Spin pumping contributions to the intrinsic damping are not significant in our films, as no heavy-metal seed layers have been used and the films have thicknesses of 10~nm or greater. For the radiative damping contribution \cite{Schoen2015} in the geometry of our CPW and sample, we calculate contributions $\alpha_{\text{rad}}\lesssim 1 \times 10^{-4}$, which is below the uncertainty in our damping fit parameter.
\begin{figure}
\includegraphics{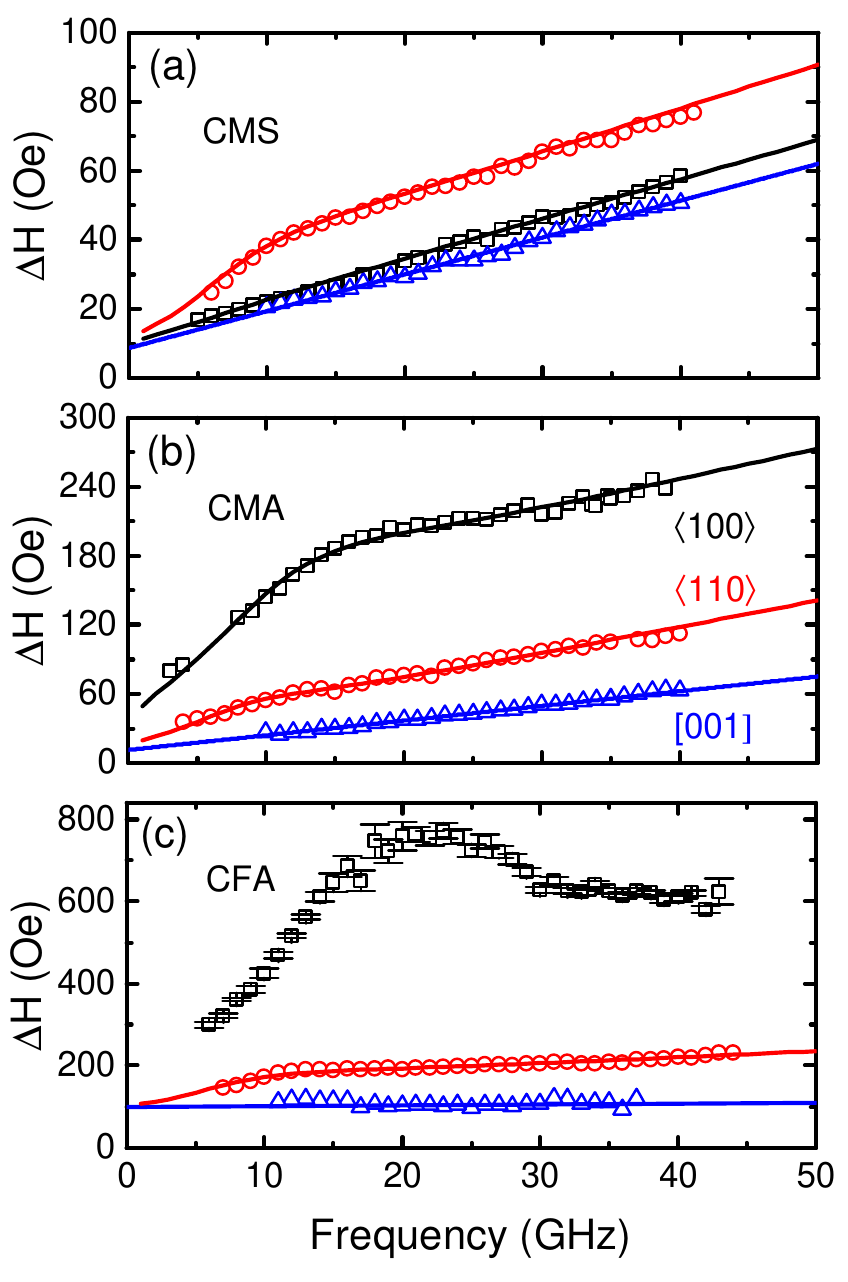}
\caption{Linewidths along all three principal directions for CMS (a), CMA (b), and CFA (c). Heusler crystalline axes are labeled by $\langle$100$\rangle$ (black), $\langle$110$\rangle$ (red), and [001] (blue). In all three cases, $\langle$110$\rangle$ is the in-plane easy axis and $\langle100\rangle$ is the in-plane hard axis. The corresponding fits are shown as the solid curves, where the in-plane linewidths are fit using Eq.\ \ref{TMS_LW} and the out-of-plane linewidths are fit to the Gilbert damping model. The fit parameters are given in Table \ref{tab:TMSparameters}.}
\label{fig:alllinewidths}
\end{figure}
\subsection{\label{IP}In-plane linewidths}
With the intrinsic damping and inhomogeneous broadening characterized by the PP measurement, we turn our attention to the IP linewidth measurements, for which TMS contributions are present. For hard-axis measurements, frequencies $\lesssim5$~GHz were not used due to the influence of slight magnetic field misalignment on the linewidths. For easy-axis measurements, the lower limit is determined by the zero-field FMR frequency. Figure \ref{fig:angledependence} shows the dependences of the resonance fields and linewidths on the angle of the in-plane field. An important observation seen in Fig.\ \ref{fig:angledependence} is that the linewidth extrema are commensurate with those of the resonance fields and therefore the magnetocrystalline anisotropy energy.  This rules out field-dragging and mosaicity contributions to the linewidth, which can occur when the resonance field depends strongly on angle \cite{Zakeri2007}. We note that similar IP angular dependence of the FMR linewidth, which was attributed to an anisotropic TMS mechanism caused by a rectangular array of misfit dislocations, has been reported by \citet{Kurebayashi2013} and Woltersdorf and Heinrich \cite{Woltersdorf2004} for epitaxial Fe/GaAs(001) ultrathin films.

To further study the anisotropy of the IP $\Delta H$ in our films, we have measured $\Delta H$ at the angles corresponding to the extrema of $H_{FMR}$ (and $\Delta H$) in Fig.\ \ref{fig:angledependence} over a range of frequencies. These data are shown in Fig.\ \ref{fig:alllinewidths}, along with the PP ([001]) measurements for each sample. A distinguishing feature of the data shown in Fig.\ \ref{fig:alllinewidths} is the significant deviation between IP and PP linewidths in all but one case (CMS$\langle100\rangle$). Large and nonlinear frequency dependence of the IP linewidths is strongly suggestive of an active TMS linewidth broadening mechanism. In the presence of TMS, careful analysis is required to separate the Gilbert damping from the TMS linewidth contributions. We therefore describe the TMS mechanism in more detail in the following section in order to analyze the IP linewidths in Fig.\ \ref{fig:alllinewidths} and extract the Gilbert damping.
\begin{figure}
    \includegraphics{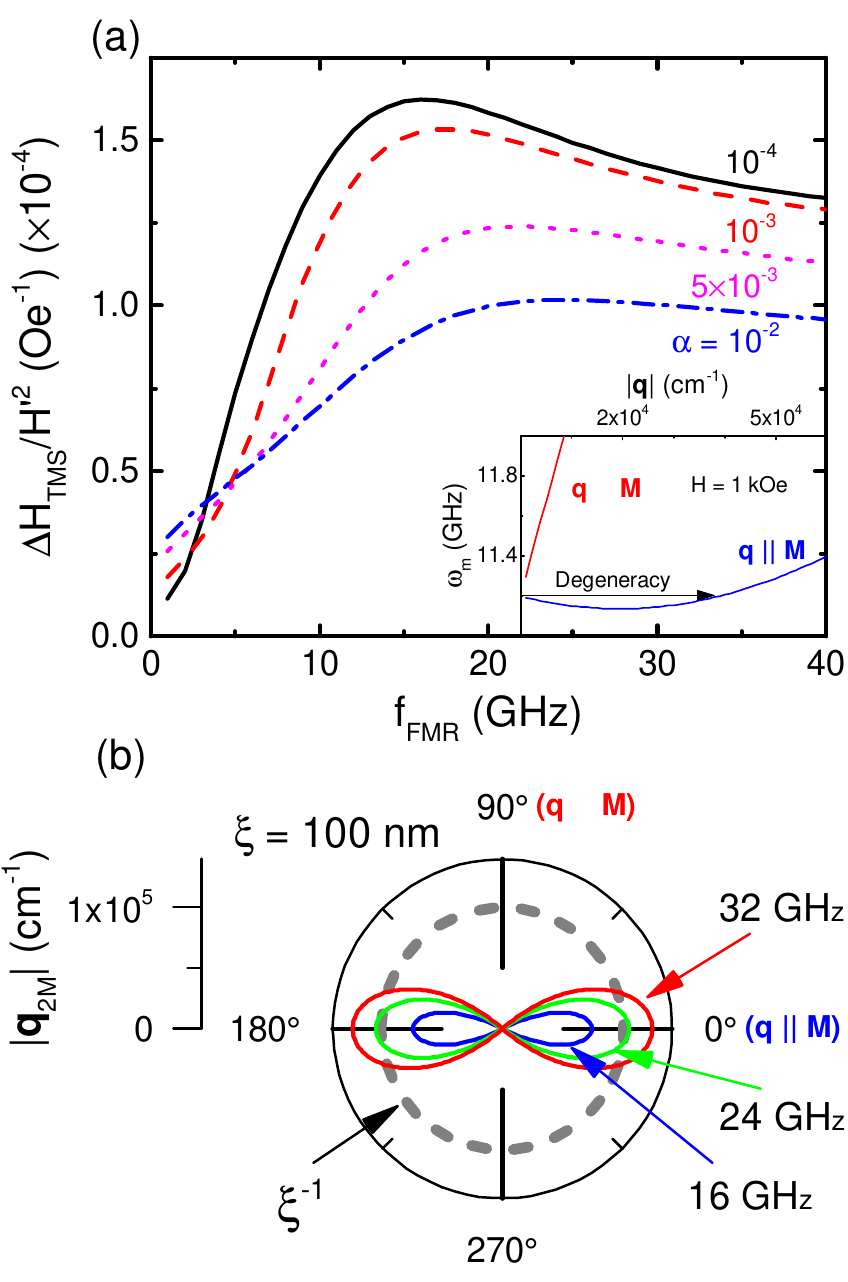}
    \caption{(a) Two-magnon scattering linewidth contribution for values of  Gilbert damping $\alpha = 10^{-2}, 5 \times 10^{-3}, 10^{-3},$ and $10^{-4}$. The inset shows magnon dispersions for an applied field of $H$~=~1~kOe. (b) Contours of the degenerate mode wavenumber $\mathbf{q}_{2M}$ in the film plane as a function of wavevector angle relative to the magnetization for $f_{FMR}$ = 16, 24, and 32~GHz. The dashed circle indicates the wavenumber of a defect with size $\xi$ = 100~nm.}
\label{fig:theorycartoon}
\end{figure}
 \subsection{Two-magnon scattering model}\label{TMS}
The TMS mechanism leads to a characteristic nonlinear frequency dependence of $\Delta H$ \cite{Arias1999,McMichael2004}. In Fig.\ \ref{fig:alllinewidths}, the IP $\Delta H$ is not a linear function of frequency, but possesses the ``knee" behavior characteristic of the frequency dependence of linewidths dominated by the TMS mechanism. We have fit our data to the TMS model described by \citet{McMichael2004}, in which the TMS linewidth $\Delta H_{TMS}$ is given by \cite{Krivosik2007,Kalarickal2008}
\begin{equation}\label{TMS_LW}
\Delta H_{TMS} = \frac{\gamma^2\xi^2H'^2}{df/dH\lvert_{f_{FMR}}}\int\Gamma_{0\mathbf{q}}C_{\mathbf{q}}(\xi)\delta_\alpha(\omega-\omega_{\mathbf{q}})d^2q,
\end{equation}
where $\Gamma_{0\mathbf{q}}$ is the defect-mediated interaction term between magnons at wavevector $0$ and $\mathbf{q}$, $C_{\mathbf{q}}(\xi) = (1+(q\xi)^2)^{-3/2}$ is the correlation function of the magnetic system with correlation length $\xi$, and $H'$ is the magnitude of the characteristic inhomogeneity (units of magnetic field). The $\delta_\alpha$-function in Eq.\ \ref{TMS_LW} selects only the magnon scattering channels that conserve energy. In the limit of zero intrinsic damping, it is identical to the Dirac delta function, but for finite $\alpha$ it is replaced by a Lorentzian function of width $\delta\omega = (2\alpha \omega/\gamma)d\omega/dH$. The magnon dispersion relation determining $\omega_{\mathbf{q}}$ is the usual Damon-Eshbach thin film result \cite{Eshbach1960,Krivosik2007} with the addition of magnetocrystalline anisotropy stiffness field terms extracted from the dependence of the resonance field on the applied frequency for the IP configuration. The film thickness $d$ affects the states available for two-magnon scattering through the dispersion relation, namely, the linear term which gives rise to negative group velocity for small $q$ ($\propto - q d$). The IP FMR linewidth data shown in Fig.\ \ref{fig:alllinewidths} were fit to Eq.\ \ref{LWtotal} (with Eq.\ \ref{TMS_LW} used to evaluate $\Delta H_{TMS}$) with $\xi$, $\alpha$, and $H'$ as fitting parameters (shown in Table \ref{tab:TMSparameters}). The correlation length $\xi$ remains approximately constant for different in-plane directions, while the strength $H'$ is larger for the $\langle100\rangle$ directions in the CMA and CFA samples and the $\langle110\rangle$ directions in the CMS sample. Some degree of uncertainty results from this fitting procedure, because for linewidth data collected over a limited frequency range, $\xi$ and $\alpha$ are not completely decoupled as fitting parameters. In absolute terms, however, the largest systematic errors come from the exchange stiffness, which is not well-known.  The error bars given in Table \ref{tab:TMSparameters} were calculated by varying the exchange stiffness over the range $400~\textrm{meV}~\textrm{\AA}^2$ to $800~\textrm{meV}~\textrm{\AA}^2$, and recording the change in the fit parameters. This range of values was chosen based on previous Brillouin light scattering measurements of the exchange stiffness in similar Heusler compounds \cite{Kubota2009,Gaier2009}. In addition, we note that in Eq.\ \ref{LWtotal} $\Delta H_0$ is taken to be isotropic, with the value given by the PP linewidth measurements shown in Fig.\ \ref{fig:PP_summary}. Although certain realizations of inhomogeneity may result in an anisotropic $\Delta H_0$ (see Ref.\ \cite{Woltersdorf2004} for a good discussion), doing so here would only serve to create an additional fitting parameter.
\subsection{Effect of low intrinsic damping}\label{alphadependence}
The effect of low intrinsic damping on the two-magnon linewidth can be seen in Fig.\ \ref{fig:theorycartoon}(a). As $\alpha$ decreases, with all other parameters fixed, $\Delta H_{TMS}$ steadily increases and becomes increasingly nonlinear (and eventually nonmonotonic) with frequency. In particular, a ``knee" in the frequency dependence becomes more pronounced for low damping (see e.g.\ Fig.\ \ref{fig:theorycartoon}(a) curve for $\alpha = 10^{-4}$). The physics giving rise to the knee behavior is illustrated in Fig.\ \ref{fig:theorycartoon}(b). The TMS process scatters magnons from zero to non-zero wavevector at small $q$.  There is assuemd to be sufficient disorder to allow for the momentum $q$ to be transferred to the magnon system.  There will always be, however, a length scale $\xi$ below which the disorder decreases, so that the film becomes effectively more uniform at large wavevectors. The corresponding FMR frequencies are those for which the contours of constant frequency (the figure eights in Fig.~\ref{fig:theorycartoon}) in $q$-space have extrema at $q \sim \xi^{-1}$. The TMS rate is also determined by the interplay of the magnon density of states, the effective area in $q$-space occupied by the modes that conserve energy, and the Gilbert damping. The knee behavior is more pronounced for low $\alpha$ due to the increased weight of the van Hove singularity coming from the tips of the figure eights, in the integrand of Eq.\ \ref{TMS_LW}. Although a larger window of energies, set by the width of $\delta_\alpha$, is available for larger $\alpha$, this smears out the singularity in the magnon density of states, removing the sharp knee in the TMS linewidth as a function of frequency. The PP measurement confirms that all of these epitaxial Heusler films lie within the range $\alpha < 2 \times 10^{-3}$. Ferromagnetic films with ultralow $\alpha$ are therefore increasingly prone to large TMS linewidths (particularly for metals with large $M_s$). The TMS linewidths will also constitute a larger fraction of the total linewidth due to a smaller contribution from the Gilbert damping. In practice, this is why experimental reports \cite{Schoen2016,Oogane2018,Guillemard2019} of ultralow $\alpha$ have almost all utilized the PP geometry.
\subsection{Discussion}\label{discussion}
\begin{table}
    \caption{Summary of the fitting parameters used to fit the in-plane data of Fig.\ \ref{fig:alllinewidths} (black squares and red circles) to Eqs.\ \ref{LWtotal} and\ \ref{TMS_LW}. CFA refers to the unannealed Co$_2$FeAl sample.}
    \label{tab:TMSparameters}
    \begin{ruledtabular}
    \begin{tabular}{l r r r}
       \\[-8pt]
       Sample (Field Direction) & $\alpha~(\times 10^{-3})$ & $\xi$ (nm) & $H^{\prime}$ (Oe) \\[4pt]
        \hline
        \\[-6pt]
        CMS$\langle$110$\rangle$ & $1.6 \pm 0.2$ & $40 \pm 25$ & $55 \pm 30$\\
        CMS$\langle100\rangle$ & $1.5 \pm 0.1$ & $40 \pm 25$ & $30 \pm 15$\\
        CMA$\langle$110$\rangle$ & $3.1 \pm 0.2$ & $70 \pm 20$ & $30 \pm 5$\\
        CMA$\langle100\rangle$ & $4.7 \pm 0.4$ & $55 \pm 10$ & $90 \pm 5$\\
        CFA$\langle$110$\rangle$ & $2.0 \pm 0.3$ & $20 \pm 10$ & $175 \pm 60$\\
        CFA$\langle100\rangle$ & N/A & N/A & N/A
    \end{tabular}
    \end{ruledtabular}
\end{table}
The results of the IP linewidth fits to Eqs.\ \ref{LWtotal} and \ref{TMS_LW} are summarized in Table \ref{tab:TMSparameters}. In the case of CMS, the high-frequency slopes in Fig.\ \ref{fig:alllinewidths}(a) approach the same value along each direction, as would be expected when the frequency is large enough for the TMS wavevector to exceed the inverse of any defect correlation length.  In this limit, $\alpha$ is isotropic (within error limits).

Next, we discuss the CMA IP data shown in Fig.\ \ref{fig:alllinewidths}(b) and Table \ref{tab:TMSparameters}. It is clear from this figure that a good fit can be obtained along both $\langle100\rangle$  and $\langle110\rangle$ directions. In Table \ref{tab:TMSparameters} it can be seen that the value of the defect correlation length $\xi$ is approximately the same along both directions. However, the values of $\alpha$ we obtain from fitting to Eqs.\ \ref{LWtotal} and \ref{TMS_LW} do not agree well with the PP value of $\alpha_{001} = 1.8 \times 10^{-3}$ (Fig.\ \ref{fig:PP_summary}). Anisotropic values of $\alpha$ have been both predicted \cite{Gilmore2010,Qu2014} and observed \cite{Chen2018}, and an anisotropic $\alpha$ is possibly the explanation of our best-fit results. The in-plane $\langle100\rangle$ and [001] directions are equivalent in the bulk, so the anisotropy would necessarily be due to an interface anisotropy energy \cite{Chen2018} or perhaps a tetragonal distortion due to strain \cite{Li2019}.

Finally, we discuss the CFA linewidths shown in Fig.\ \ref{fig:alllinewidths}(c) and Table \ref{tab:TMSparameters}. This sample has by far the largest two-magnon scattering contribution, which is likely related to the anomalously large inhomogeneous broadening and low intrinsic damping [see Fig.\ \ref{fig:theorycartoon}(a)] observed in the PP measurement. A good fit of the data was obtained when the field was applied along the $\langle110\rangle$ direction. Notably, the IP $\langle110\rangle$ best fit value of $2.1\times 10^{-3}$ is nearly a factor of 3 larger than the $\alpha_{001}$ upper bound on the same sample (Table \ref{tab:magneticparameters}), strongly suggesting an anisotropic Gilbert $\alpha$. A striking anisotropy in the IP linewidth was revealed upon rotating the magnetization to the $\langle100\rangle$ orientation. For the $\langle100\rangle$ case, which yielded the largest TMS linewidths measured in this family of films, we were not able to fit the data to Eq.\ \ref{TMS_LW} using a set of physically reasonable input parameters. We believe that this is related to the consideration that higher order terms in the inhomogeneous magnetic energy (see Ref.\ \cite{Krivosik2007}) need to be taken into account. Another reason why this may be the case is that the model of \citet{McMichael2004} assumes the inhomogeneities to be grain-like, whereas the samples are epitaxial [see Fig.\ \ref{fig:1}(a)]. Atomic force microscopy images of these samples [Fig.\ \ref{fig:1}(d)] imply that grains, if they exist, are much larger than the defect correlation lengths listed in Table \ref{tab:TMSparameters}, which are of order 10's of nm. We also note that there does not appear to be a correlation between the strength of two-magnon scattering $H'$ and the cubic anisotropy field $2K_1/M_s$, which would be expected for grain-induced two-magnon scattering.
\section{Summary and Conclusion}\label{summary}
We conclude by discussing the successes and limitations of the \citet{McMichael2004} model in analyzing our epitaxial Heusler film FMR linewidth data. We have shown that two-magnon scattering is the extrinsic linewidth-broadening mechanism in our samples. Any model which takes this as its starting point will predict much of the qualitative behavior we observe, such as the knee in the frequency dependence and the large linewidths IP for low $\alpha$ films. The TMS model used in this article (for the purpose of separating TMS and Gilbert linewidth contributions) is, however, only as accurate as its representation of the inhomogeneous magnetic field and the underlying assumption for the functional form of $C_{\mathbf{q}}(\xi)$. Grain-like defects are assumed, which essentially give a random magnetocrystalline anisotropy field. We did not, however, explicitly observe grains in our samples with AFM, at least below lengthscales of $\sim$10~$\mu$m [Fig.\ \ref{fig:1}(d)]. Misfit dislocations, a much more likely candidate in our opinion, would cause an effective inhomogeneous magnetic field which could have a more complicated spatial profile and therefore lead to anisotropic two-magnon scattering (see Ref.\ \cite{Woltersdorf2004}). The perturbative nature of the model also brings its own limitations, and we believe that the CFA$\langle100\rangle$ data, for which we cannot obtain a satisfactory fit, are exemplary of a breakdown in the model for strong TMS. Future work should go into methods of treating the two-magnon scattering differently based on the type of crystalline defects present, which will in turn allow for a more reliable extraction of the Gilbert damping $\alpha$ and facilitate the observation of anisotropic Gilbert damping, enabling quantitative comparison to first-principles calculations.

Regardless of the limitations of the model, we emphasize three critical observations drawn from the linewidth measurements presented in this article. First, in all cases we observe large and anisotropic TMS linewidth contributions, which imply inhomogeneity correlation lengthscales of order tens-to-hundreds of nanometers. The microscopic origin of these inhomogeneities is the subject of ongoing work, but are likely caused by arrays of misfit dislocations \cite{Woltersdorf2004}. The relatively large lengthscale of these defects may cause them to be easily overlooked in epitaxial film characterization techniques such as XRD and cross-sectional HAADF-STEM, but they still strongly influence magnetization dynamics. These defects and their influence on the FMR linewidth through TMS complicate direct observation of Kambersk\'y's model for anisotropic and (in the case of Heusler compounds) ultralow intrinsic damping in metallic ferromagnets. Second, we observed low intrinsic damping through our PP measurement, which was $<2\times10^{-3}$ for all of our samples. Finally, we have presented the mechanism by which FMR linewidths in ultralow damping films are particularly likely to be enhanced by TMS, the anisotropy of which may dominate any underlying anisotropic Kambersk\'y damping.

This work was supported by NSF under DMR-1708287 and by SMART, a center funded by nCORE, a Semiconductor Research Corporation program sponsored by NIST. The sample growth was supported by the DOE under DE-SC0014388 and the development of the growth process by the Vannevar Bush Faculty Fellowship (ONR N00014-15-1-2845). Parts of this work were carried out in the Characterization Facility, University of Minnesota, which receives partial support from NSF through the MRSEC program.
\end{document}